\documentclass[letterpaper,aps,prb,twocolumn,superscriptaddress,floatfix]%
{revtex4}

\usepackage[dvips]{color}
\usepackage{graphicx}
\usepackage{natbib}
\usepackage[utf8]{inputenc}

\begin{document}


\title{Analysis of the spectral function of Nd$_{1.85}$Ce$_{0.15}$CuO$_4$,
obtained by angle resolved photoemission spectroscopy}


\author{F. Schmitt$^1$, W. S. Lee$^1$, D.-H. Lu$^2$, W. Meevasana$^3$, E.
Motoyama$^1$, M. Greven$^{1,2}$, Z.-X. Shen$^{1,2,3}$}
\affiliation{%
$^{1}$Department of Applied Physics, Stanford University, Stanford, CA 94305,
USA\\
$^{2}$Stanford Synchrotron Radiation Laboratory, Menlo Park, CA 94025, USA\\
$^{3}$Department of Physics, Stanford University, Stanford, CA 94305, USA}

\date{\today}

\begin{abstract}

Samples of Nd$_{2-x}$Ce$_x$CuO$_4$, an electron-doped high temperature
superconducting cuprate (HTSC), near optimal doping at $x=0.155$ were measured
via angle resolved photoemission (ARPES). We report a renormalization feature in
the self energy (``kink'') in the band dispersion at $\approx 50-60$ meV present
in nodal and antinodal cuts across the Fermi surface. Specifically, while the
kink had been seen in the antinodal region, it is now observed also in the nodal
region, reminiscent of what has been observed in hole-doped cuprates.
\end{abstract}


\maketitle


The high temperature superconducting cuprates (HTSCs) have received much
attention since their discovery, as the mechanism of their unusually high
critical temperatures remains yet to be determined. Nd$_{2-x}$Ce$_x$CuO$_4$
(NCCO) is a class of cuprates that resides on the relatively less-studied
electron (n)-doped side of the phase diagram\cite{Tokura1989}, which is
qualitatively different from the hole (p)-doped side. NCCO has recently been
attracting increased interest\cite{Matsui2005, Armitage2001_2,
Armitage2002, Armitage2003, Yamada2003, Motoyama2007, Li2006, Dagan2004}, in
particular regarding its links to their more commonly studied hole-doped
counterparts \cite{Damascelli2003}.

Angle-resolved photoemission spectroscopy (ARPES) is used as a powerful direct
probe of the electronic band structure (the one-particle spectral function)
\cite{Damascelli2003}. Sudden changes in the slope in the band dispersion
observed via ARPES allow one to infer the underlying physics causing these
renormalizations \cite{Damascelli2003}.
Low-energy renormalizations have been prevalently found in the nodal
region of p-doped cuprates via ARPES\cite{Damascelli2003, Lanzara2004}, but not
in the nodal region of n-doped cuprates \cite{Armitage2003}, casting doubt on
the universality of the renormalization effect in the cuprates for the whole
phase diagram.

In this communication, we confirm a renormalization in NCCO near optimal
doping ($x=0.155$) around 55~meV in the antinodal (X-M) region and report one in
the nodal ($\Gamma$-M) region of about the same energy which was not observed
earlier \cite{Armitage2003}. We do not observe any change of these
renormalizations across the superconducting (SC) phase transition. These
results suggest that an oxygen phonon mode with comparable energy scale is a
likely origin of these renormalizations. In light of the presence of phononic
coupling effects, one needs to carefully exam the data to separate the
respective contributions to the low energy spectra from AF band folding and
lattice coupling effects, as the latter can also break the low energy dispersion
into two ``branches''.

ARPES data were taken at beamline 5-4 of the Stanford Synchrotron Radiation
Laboratory with Scienta SES200 and R4000 analyzers at a photon energy of
16.75~eV. The a-(b-)axis was aligned at $45^\circ$ to the light polarization for
the nodal cut, and parallel to the light polarization for the other cuts. The
energy resolution was $\approx$ 10~meV and the angular resolution $\approx$
0.3$^\circ$. All samples were cleaved at pressures better than
$3\times10^{-11}$~torr; the measurement temperature was 10~K unless otherwise
specified. Single crystals of NCCO were grown at Stanford
University\cite{Motoyama2006}. The doping level, determined by inductively
coupled plasma spectroscopy, is $15.5\%\pm0.7\%$ Ce. $T_C$ was determined by
SQUID magnetometry to be 25~K with a transition width of $\approx$2~K.

\begin{figure}[!t]
\includegraphics[width=0.45\textwidth]{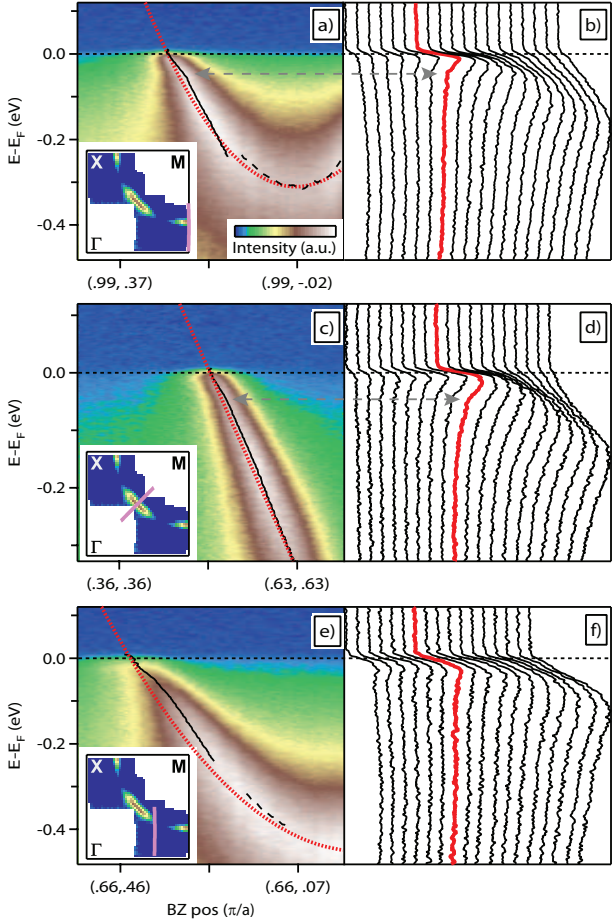}%
\caption{(Color online) FS cuts and EDCs of the antinodal (a, b), nodal (c, d),
and hot spot (e, f) regions. The insets in a), c), and e) depict a map of the
upper right quadrant of the FS; the cut positions are indicated by pink (gray)
lines. a), c), and e) show the raw spectra, while b), d), and f) show EDCs near
the Fermi-crossing of the respective raw spectra on the left. Also indicated in
the raw spectra are the MDC fits in solid black, EDC fits in dashed black, and
assumed bare band in dotted red (black). The EDCs in b), d), and f) at $E_F$
crossing are highlighted in thick red (black). A renormalization is seen in the
nodal and antinodal regions around 50 meV (gray dashed arrows).}
\label{fig:raw_cuts_with_EDCs}
\end{figure}

We show three different cuts through the Fermi surface (FS) in fig.
\ref{fig:raw_cuts_with_EDCs}: one in the antinodal, one in the nodal, and one
near the hot spot region (region of low intensity at $E_F$ near the crossing of
the FS with the antiferromagnetic (AF) Brillouin zone (BZ) boundary). The insets
in fig. \ref{fig:raw_cuts_with_EDCs} a, c, and e show the respective locations
of the three cuts in the BZ relative to the FS of NCCO. The raw data of these
three cuts, including their EDC curves, are shown in fig.
\ref{fig:raw_cuts_with_EDCs} a-f. MDC analysis, i.e. a fit with a Lorentzian
lineshape, is used to extract Re $\Sigma$ and Im $\Sigma$: Im $\Sigma$
corresponds to the extracted width of the Lorentzian times the Fermi velocity,
while Re $\Sigma$ is obtained by subtracting the extracted position from an
assumed bare band. Apart from the assumptions and details of this
analysis\cite{Armitage2003, Kaminski2001}, we phenomenologically apply MDC/EDC
spectral function analysis as a tool for quantification, lacking a definitive
theory.

The raw spectrum in fig. \ref{fig:raw_cuts_with_EDCs} a of the antinodal cut
reveals a renormalization at about 50~meV binding energy. The corresponding EDC
curves in fig. \ref{fig:raw_cuts_with_EDCs} b display a sharp quasiparticle peak
near the $E_F$ crossing. This peak terminates with a dip at about 50~meV binding
energy. The linewidth decreases upon approaching $E_F$. To determine the self
energy, a parabolic bareband has been assumed and constructed by fixing the
$E_F$-crossing point and the band bottom, which are determined by a linear fit
to a MDC analysis $<40$ meV, and a parabolic fit to an EDC analysis $\geq 250$
meV, respectively (cf. fig. \ref{fig:raw_cuts_with_EDCs} a). The band bottom
thus obtained is at about $300$ meV. The discrepancy between the EDC and MDC
fitting around $250$ meV shows the limits of both
analyses \cite{Damascelli2003}. 

The raw spectrum of the nodal region cut depicted in fig.
\ref{fig:raw_cuts_with_EDCs} c exhibits a similar renormalization feature at a
similar energy of about 50~meV as seen in the antinodal region. The EDCs (fig.
\ref{fig:raw_cuts_with_EDCs} d) show a peak at the $E_F$ crossing. Again, the
peak terminates at roughly 50~meV. A resemblance is found in the peak-dip-hump
(PDH) structure seen over a wide temperature range in the nodal EDCs of the
single-layer compound Bi2201 \cite{Lanzara2004} which is attributed to
electron-phonon interaction, although a hump is absent here in the antinodal and
faint in the nodal cuts. The assumed linear bare band was modeled by connecting
the $E_F$ crossing point obtained by a linear fit to the MDC dispersion
$< 40$~meV, and a point at 300~meV obtained by a linear fit to the MDC
dispersion between 250 and 330~meV (cf. fig. \ref{fig:raw_cuts_with_EDCs} c). We
note that assuming different bare bands changes the extracted renormalization
strength, but not appreciably the energy of sharp features in the real part of
the self energy. 


\begin{figure}[!b]
\includegraphics[width=0.45\textwidth]{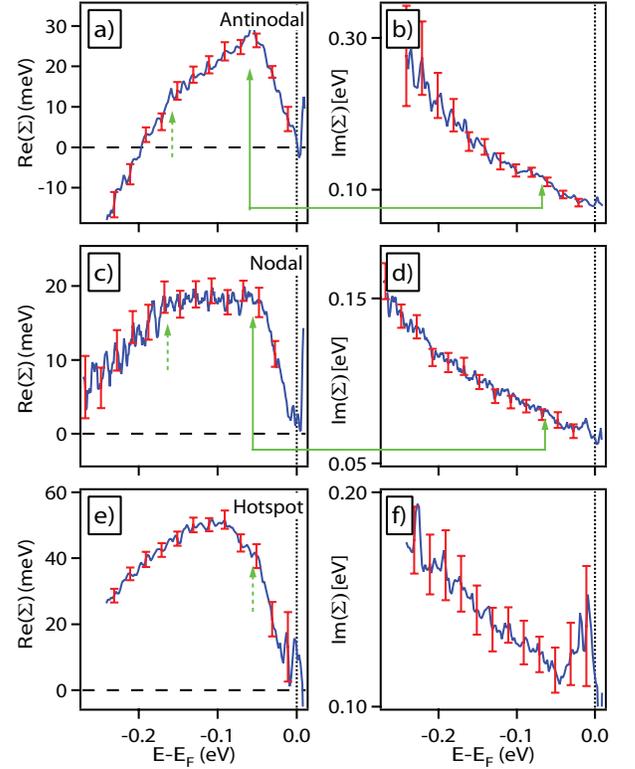}%
\caption{(Color online) Self energies extracted from the antinodal (a, b), nodal
(c, d), near-hot spot (e, f) cuts shown in fig. \ref{fig:raw_cuts_with_EDCs},
respectively. a), c), and e) show the respective real part; b), d), and f) the
respective imaginary part. Changes in Re $\Sigma$ and their corresponding drops
in Im $\Sigma$ are marked by green (gray) arrows. The error bars at selected
points represent the $3 \sigma$ confidence levels from the MDC-Lorentzian fit
position in the Re $\Sigma$ graphs, and the MDC-Lorentzian fit HWHM in the Im
$\Sigma$ graphs. The dashed arrows mark weak changes in Re $\Sigma$ that are
hard to distinguish from the statistical background.}
\label{fig:real_im_selfEs}
\end{figure} 

The spectrum of the hot spot region is displayed in fig.
\ref{fig:raw_cuts_with_EDCs} e. Since the hot spot is located on the AF BZ
boundary, the low-energy physics and thus the kink will be influenced by a
crossover of two bands that could originate from backfolding at the AF BZ
boundary as we discuss later. Bare band assumption and construction are
analogous to the antinodal region.

The real and imaginary parts of the self energy are displayed in fig.
\ref{fig:real_im_selfEs} a-f. A peak in Re $\Sigma$ is clearly visible at
$60\pm5$ meV in the antinodal region (fig. \ref{fig:real_im_selfEs} a), with a
corresponding drop at about the same energy in Im $\Sigma$ (fig.
\ref{fig:real_im_selfEs}b), as expected from causality arguments. Likewise, in
the nodal region a plateau is observed at about $50\pm10$ meV in Re $\Sigma$
(fig. \ref{fig:real_im_selfEs} c). The drop in Im $\Sigma$ (fig.
\ref{fig:real_im_selfEs} d) is much less clear, but the data seem qualitatively
consistent with the Kramers-Kronig relations. The difference in our ability to
extract the real and imaginary parts of the self-energy can easily be understood
from the fact that, for a relatively broad feature, it is much easier to
determine its position than width. 

We believe the reason why we are now able to resolve the kink in the nodal
region and not before\cite{Armitage2003} is mainly due to the improvement of the
momentum resolution. The photon energy (16.75~eV) we used as compared to that
used in previous measurements(53~eV) doubles our resolution in k-space. Indeed,
the line width --- which depends significantly on sample quality and resolution
--- is smaller (cf. fig. \ref{fig:real_im_selfEs} b, d) in our data presented in
this communication. 

\begin{figure}[!t]
 \includegraphics[width=.45\textwidth]{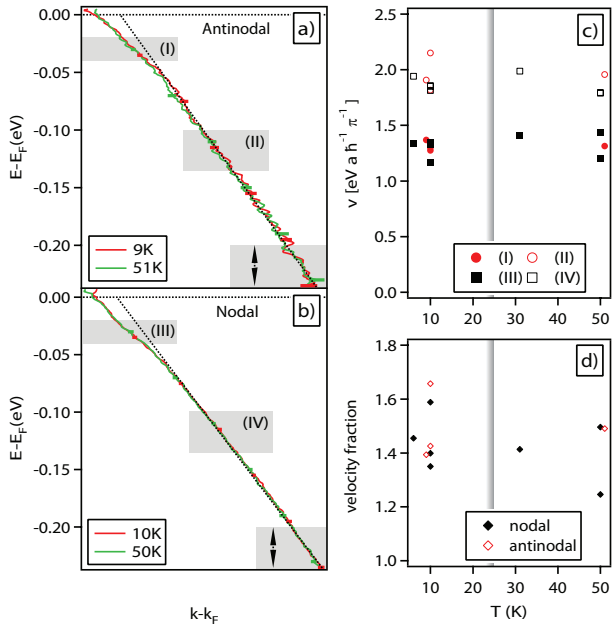}%
\caption{(Color online) Temperature dependence of the MDC-derived dispersions in
the antinodal (a) and nodal (b) regions, with 3$\sigma$ errors from the
fit (thick horizontal bars). The black dotted lines serve as guide to the eye.
The data have been average-subtracted in the gray region (dotted arrow) to
overlay them. c) Temperature dependent velocities in both nodal and antinodal
region above and below the kink energy, extracted from gray regions indicated by
I-IV in a) and b). d) Temperature dependent fractions of the respective
velocities below and above the kink energy for the nodal and antinodal region.
$T_C$ is indicated in c) and d) (vertical gray line).}
\label{fig:temp_dep}
\end{figure} 

A change in slope can also be seen in Re $\Sigma$ near the hot spot region (fig.
\ref{fig:real_im_selfEs} e) around $50\pm10$ meV. However, the relation of its
cause to the features in the nodal and antinodal regions remains speculative,
even though we find its appearance at the same energy suggestive.

Lastly, a weak change of slope at a higher energy of around $160$ meV can be
observed in Re $\Sigma$ in both the nodal and antinodal region (fig.
\ref{fig:real_im_selfEs} a,c). Within the statistics of the data, no change is
visible in either of the corresponding Im $\Sigma$ (fig.
\ref{fig:real_im_selfEs} b,d) at this energy. It is currently being investigated
and will not be discussed here.

The temperature dependence of the observed ubiquitous feature in the nodal and
the antinodal region is shown in fig. 3. Within the statistics, there is no
change visible in either region below ($<$6~K, 10~K) and above (30~K, 50~K)
the SC transition.

For further clarification, the velocities obtained by linear fits against the
MDC Lorentz position above (-40~meV$<E-E_F<$-20~meV) and below
(-135~meV$<E-E_F<$-100~meV) the renormalization are graphed versus temperature
for both nodal and antinodal regions in fig. \ref{fig:temp_dep} c. The data
result from seven different measurements on four samples in the nodal, and four
different measurements on three samples in the antinodal region. First, neither
of these velocities change appreciably across the SC transition. Second, the
Fermi velocities of the nodal and antinodal regions are the same, confirming
previous results\cite{Armitage2003}. Third, the ratio of renormalized and
unrenormalized velocities --- which are indicative of the renormalization
strength of the electronic band --- do not seem to change within the statistics
in both nodal and antinodal regions, cf. fig. \ref{fig:temp_dep} d.

\begin{figure}[!t]
 \includegraphics[width=0.45\textwidth]{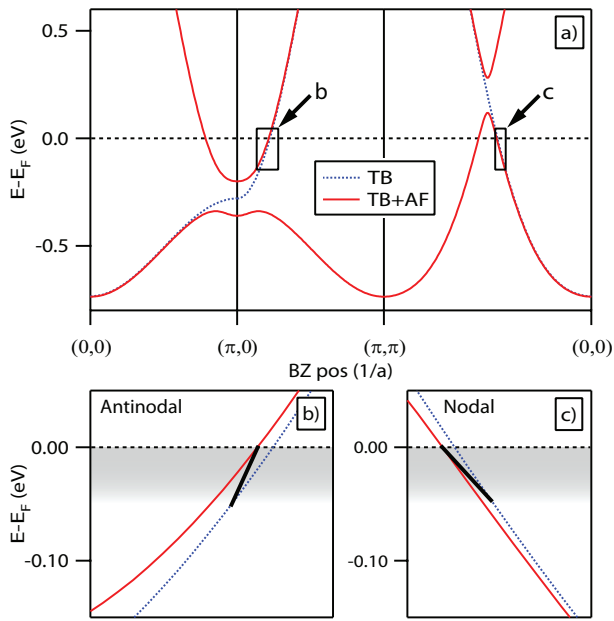}
\caption{(Color online) Schematic tight binding (TB) models. Solid blue (black)
curves correspond to TB without, dotted red (black) ones to TB with simple
perturbative $V_{\pi,\pi}$ scattering. a) cuts along high symmetry directions in
the BZ. b) antinodal and c) nodal details of a). The enlarged regions shown in
b) and c) are indicated by rectangles in a).}
\label{fig:bandfoldingschema}
\end{figure}

The n-doped HTSCs possess hot spots that lie on the AF BZ
boundary\cite{Armitage2001_2, Armitage2003}, and recently, pockets were observed
suggesting a band back-folding at the AF BZ boundary\cite{Park2007}; a possible
mechanism could be $(\pi,\pi)$-scattering due to some short range or remnant
antiferromagnetic (SDW) ordering. To first order, this folding results in a pair
of new band sheets (cf. fig. \ref{fig:bandfoldingschema}
a)\cite{Armitage2001_3}. Whenever there is a crossover from one folded band
sheet to another, as for example at $(\pi,0)$ in the antinodal region (solid
curve to solid curve in fig. \ref{fig:bandfoldingschema}), one expects a
kink-like feature in the dispersion. We do not however think that this is the
sole cause for the low energy feature, since this crossover takes place at the
band bottom ($\approx$ 300~meV) in the antinodal and above the Fermi energy in
the nodal region, leading to a highly anisotropic behavior both in shape and
energy of the feature, contradicting our observed isotropy in both energy and
renormalization strength. 

If we were to invoke an electron energy dependence of the scattering,
$V_{\pi,\pi}(\omega_k)$, whereas the interaction is nonzero only for $|\omega_k|
< \omega_{max} \approx 50$~meV, a fade-over is produced between the
unrenormalized, unfolded band (dotted curve in fig. \ref{fig:bandfoldingschema})
and one of the renormalized, folded bands (solid curves; cf. fig.
\ref{fig:bandfoldingschema} b and c) around $\omega_{max}$. 
However, we do not know of any interaction with above-mentioned energy
dependence. Provided it exists, the produced crossover results in a downwards
kink in the nodal and an \emph{upwards} kink in the antinodal region,
contradicting our observations.

We also argue against a magnetic resonance or spin flip waves as the origin of
the low energy renormalization we discuss here, since the characteristic energy
scales for electron-doped cuprates are too different from the 55~meV observed
by us. First, the relevant region of the spin excitation dispersion
relation is about 0.3~eV \cite{Wilson2006_2}. Second, a resonance has been
reported at about 10~meV in PLCCO \cite{Wilson2006} and in NCCO \cite{Zhao2007}.
This is still disputed; at least for NCCO, recent data show even lower energy
scales (within 4-8~meV) for the SC gap and a possible resonance
\cite{Yu2008}. Whether a more complicated interaction mechanism could explain
our data is to be investigated.

A phononic origin on the other hand could account for the isotropy of the
renormalization strength and energy of the kink observed by us in a simple
picture (although more intricate phononic interactions may result in anisotropic
renormalizations \cite{Shen2002, Devereaux2004} and even d-wave SC gaps
\cite{Newns2007, Shen2002}). Furthermore, a softening with
doping of oxygen phonon modes in the relevant energy regime has been observed
via inelastic x-ray scattering \cite{dAstuto2002} and inelastic neutron
scattering \cite{Kang2002}. Raman and IR spectra also show the existence of
oxygen modes of the relevant energies \cite{Litvinchuk1991}. In addition, the
fact that the kink at both nodal and antinodal region does not change across the
SC transition lends further support to the electron-phonon coupling scenario.
The absence of a change of the kink energy across the superconducting phase
transition is due to the small superconducting gap in NCCO.  Also, the EDCs at
the $E_F$ crossing exhibit the same structure as in Bi2201 --- a small peak near
$E_F$ and a shallow dip at the same energy of the renormalization seen in Re
$\Sigma$ (cf. fig. \ref{fig:raw_cuts_with_EDCs} b, d) \cite{Lanzara2004}.

Within the context of electron-phonon interaction, a renormalization at around
70~meV is seen in the nodal region of p-doped HTSCs in ARPES and has been
attributed to electron-phonon interactions\cite{Lanzara2001, Damascelli2003}. A
softening of an oxygen mode was seen as well for the n-doped compound NCCO via
inelastic neutron and X-ray scattering (55~meV)\cite{Braden2005, dAstuto2002},
and now also in the nodal region via ARPES as observed by us in the same
relevant energy region. Independently, Park et al.\cite{Park2008} and Liu et
al.\cite{Liu2008} discovered a kink in the nodal region in several n-doped HTSCs
and came to very similar conclusions regarding its origin. \footnote{We
can not, however, currently confirm the feature at $\approx$23~meV in the nodal
region observed by Liu et al.}. If interpreted as due to electron-phonon
coupling, our results suggest that electronic coupling to the oxygen phonon mode
may be even more universal, now stretching across the phase diagram. We
speculate that these phonon modes may play a vital role in determining the low
energy physics of the HTSCs \cite{Shen2002, Newns2007}.

\begin{acknowledgments}
The authors are indebted to C. Kim, P. Armitage, T. Devereaux, B. Moritz, and S.
Johnston for enlightening discussions. The SSRL/Stanford work was supported by
DOE Office of Science, Division of Materials Science, with contracts
DE-FG03-01ER45929-A001 and DE-AC02-76SF00515, and NSF grants DMR-0604701 and
DMR-0705086.
\end{acknowledgments}

\bibliography{text}

\end{document}